\documentstyle [epsfig] {l-aa}

\newif\iffigure

\begin{document}

\thesaurus{  }
 
\title{Detection of WR stars in the metal--poor starburst galaxy IZw 18}
%  \subtitle{}

\author{F. Legrand \inst{1}, D. Kunth \inst{1} , J.-R. Roy \inst{2} , J.M. Mas-Hesse \inst{3}, 
and J.R. Walsh \inst{4}\thanks{DK and JRR Visiting astronomers at  
Canada-France-Hawaii Telescope, which is operated by the National Research Council of Canada,
the Centre National de la Recherche Scientifique de France,
and the University of Hawaii} }
\offprints{F. Legrand}

\institute{
  Institut d'Astrophysique de Paris, CNRS, 98bis boulevard Arago, F-75014 
  Paris, France.
  \and
  D\'epartement de physique and Observatoire
      du mont M\'egantic, Universit\'e Laval, Qu\'ebec
      Qc G1K 7P4
  \and
  LAEFF, Apdo 50727, E-28080 Madrid,
  Spain.
  \and
  European Southern Observatory, Karl-Schwarzschild-Str. 2, 
  D-85748 Garching, 
  Germany 
}

\date{received 26 Jun 1997 ; accepted 10 Jul 1997}

\maketitle
\markboth{ F. Legrand et al.: Detection of WR stars in IZw18}{  }
\begin{abstract}

Wolf-Rayet stars (WR)  have been detected in the NW region of the metal--poor
 starburst galaxy IZw 18. The
integrated luminosity and FWHM of the bumps at 4650 \AA \ and 5808 \AA \ are
 consistent 
with the presence of a few individual stars of WC4 or WC5 type. Evolutionary
 synthesis models
predict few WRs in this galaxy, but only of WN type. 
The presence of WC stars at such low metallicity could however be explained by 
high mass loss rates, which would constrain the IMF upper mass cut-off in
IZw 18 to be higher than 80 M$_\odot$ or alternatively favor a binary channel
for WR formation.  
WC stars could also explain the strong and
narrow HeII 4686\AA \ emission line which peaks co--spatially
with the WR bump emission, as suggested by Schaerer (1996).
This detection shows that WR stars, even of WC type, are formed at
metallicities below 1/40th solar.

\end{abstract}
 
% +++++++++++++++++++++++++++++++++++++++++++++++++++++++++++++++++++++++

\section{ Introduction }

 IZw 18 is known 
to be the most metal deficient object among the blue compact dwarf galaxies
 (BCDs), with a metallicity of 1/40th of the solar 
value  and undergoing a strong star formation event (Searle \& Sargent 1972; Skillman \& Kennicutt
 1993, hereafter SK93). Moreover, IZw 18 is a 
close by object  with a recession velocity of 
$\rm 740 \ \pm \ 10 \ km/s$. 
This makes this galaxy an excellent laboratory for studying the properties of star formation at low 
metallicity. It is well known that the spectrum of IZw 18 presents a strong $\rm HeII4686\AA$ narrow 
emission line. As the ionising spectra of ordinary O stars are unable to explain
 the presence of this feature, Bergeron (1977) originally proposed
that this line can  directly originate in the atmosphere of hot Of stars.\\
Broad WR features are often found in the spectra of starburst 
galaxies (Vacca \& Conti 1992).
 As the WR stage occurs after a few Myrs in the 
lifetime of massive stars, starburst galaxies are often dominated by a recent burst of star formation 
undergoing a WR--rich evolutionary phase (Schaerer \& Vacca 1996, hereafter SV96).\\
 However, metallicity is a crucial 
parameter
for the evolution of massive stars through the WR phase in a starburst (Maeder \& Meynet 1994; 
Cervi\~no \& Mas-Hesse 1994, hereafter CMH94; Meynet 1995, hereafter M95). Specifically, when the 
metallicity decreases, the time duration 
of the WR stage decreases and the lower mass limit for a star to be able to evolve to  WR
phase increases. 
This results in a dramatic diminution of the WR/O star ratio with metallicity. Moreover, as the WC 
star progenitors 
are supposed to be more massive than the WN ones, the ratio WC/WN should 
also decrease with metallicity (M95). 
At low metallicity however, evolutionary models predict that WN stars must dominate the WR population 
(M95). At the metallicity of IZw 18, no WC 
should be formed (CMH94). \\
In Section II, we will present the observations and the measurements. 
Contrary to expectation, evidence for the presence of few WC stars will be given;
 the possible excitation of the narrow HeII line
by these stars and comparison with the evolutionary models are discussed in the last
 Section.

% +++++++++++++++++++++++++++++++++++++++++++++++++++++++++++++++++++++++

\section{ Observations and data analysis}

%\section{ Observations and reduction}

Seventeen exposures of 3000 seconds each of the blue compact galaxy IZw 18 
were obtained with the 3.6m CFH telescope during the three successive nights 
between 1995 February 1st and 4th using the MOS spectrograph 
with the 2048x2088 Loral 3 CCD detector. 
A long slit (1.52 arcsec wide) was used with a position angle 
of 45$\degr$, covering a spectral range from 3700 to 6900 $\rm \AA$. 
The slit was centered on the  central HII knot of the NW region of IZw 18. 
The spatial resolution was 0.3145 arcsec/pix and the dispersion 
$\rm 1.58 \ \AA /pix$ giving a spectral resolution of about $\rm 8.2 \ \AA$. 
The seeing was between 1 and 1.5 arcsec. The spectra were reduced using IRAF. 
Due to a slight offset between the first night and the following (less than 1''),
the sampled spatial region is slightly increased with respect to the slit width. \\

The strong emission lines in the integrated spectrum were measured
over 25 pix (7.8'') centered on the continuum maximum emission. 
This allowed us to determine a reddening of  E(B-V)=0.1, in agreement 
with SK93, assuming an 
underlying Balmer stellar absorption of 2 \AA \ EW. We have used this value to
correct the measured flux from the reddening effect. The EW of $H\beta$ is
 measured to be  $70 \ \pm \ 5$ \AA. The line measurements  are given 
in table \ref{tab:mesures}.  \\

%.................................................. begin table
    \begin{table}
    \scriptsize
    \caption [] {Relative emission line fluxes in  IZw 18 measured in the  7.8'' long integrated
spectrum.The Balmer line fluxes have been corrected for underlying stellar absorption
 by 2 \AA \ of equivalent width.} 
   \label{tab:mesures}
    \begin{flushleft}
    \begin{tabular}{cccc}
    \hline
Lines & $I(\lambda)/I(H\beta)$ & $I_{0}(\lambda)/I(H\beta)$ & 
$I_{th}(\lambda)/I(H\beta)$  \\
\hline
\ [OII]$\lambda$3727     &   0.356	&	0.388	&		0.845	\\
\ [NeIII]$\lambda$3867   &   0.181	&	0.193	&		0.176	\\
\ H$\delta$            &   0.267	&	0.283	&		-----	\\
\ H$\gamma$            &   0.476	&	0.497	&		-----	\\
\ [OIII]$\lambda$4363    &   0.064	&	0.067	&		0.049	\\
\ HeI$\lambda$4471     &   0.022	&	0.022	&		0.031	\\
\ H$\beta$             &   1.000	&	1.000	&		1.000	\\
\ [OIII]$\lambda$5007    &   2.002	&	1.979	&		1.960	\\
\ HeI$\lambda$5876     &   0.063	&	0.058	&		0.082	\\
\ H$\alpha$            &   2.978	&	2.669	&		2.830	\\
\ [NII]$\lambda$6583     &   0.008	&	0.007	&		0.009	\\
\ [SII]$\lambda$6716     &   0.021	&	0.019	&		0.007	\\
\ [SII]$\lambda$6731     &   0.016	&	0.014	&		0.005	\\
    \hline						
\ WRbump region & 	& 		&				\\
    \hline						
\ CIII4650	     &	-----   &	-----	&	    2.66E-7		\\
\ OII4651	     &	-----   &	-----	&	    1.8E-4		\\
\ [FeIII]4658	     &	-----   &	-----	&	    1.33E-4		\\
\ HeII4686	     &	0.040	&	0.041	&	    1.9E-9		\\
\ [ArIV]4713	     &	0.01    &	0.01 	&	    9.54E-3		\\
\ [ArIV]4741	     &	5.7E-3  &	5.7E-3	&	    6.77E-3		\\
    \hline
    \end{tabular}
    \end{flushleft}
{\em Columns numbers: \\}
2: Measured flux relative to H$\beta$. \\
3: Flux corrected for the reddening. \\
4: Flux predicted from SL96. \\
   \end{table}
%.................................................. end table

The most striking aspect of this spectrum, integrated  over 25 pix (7.8''),
are  two faint broad emission 
features around 4650 \AA \ and  near 5812 \AA  \ (Fig.\ref{fig:wrbumps}).

\psfig{figure=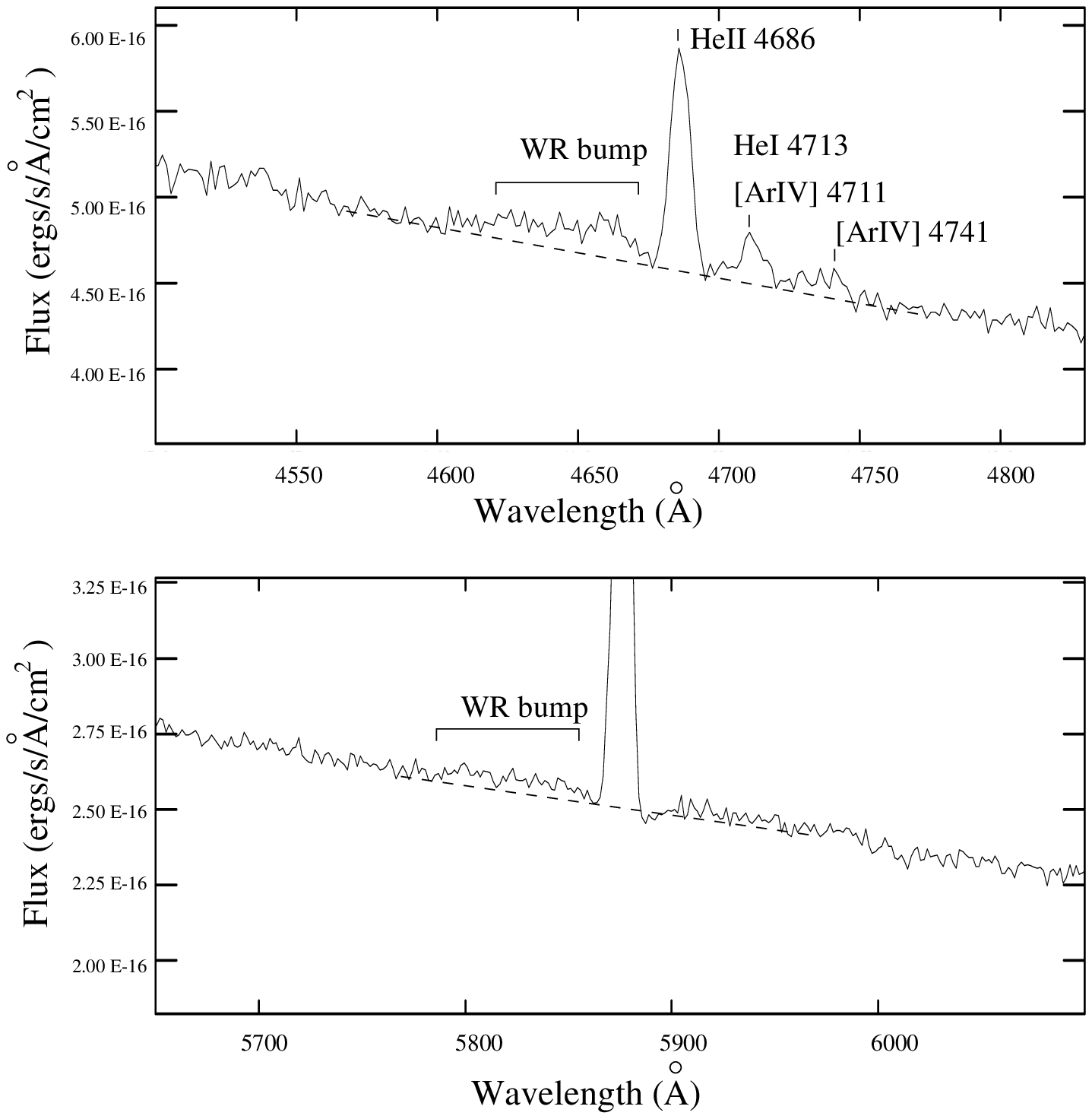,height=7.5cm,angle=-90}
\begin{figure}
%\picplace{7.5cm}
\caption[ ]{Regions of the spectrum of IZw 18 around the HeII4686\AA\ (upper) 
line and the HeI5876\AA\ (lower). 
The spectrum is integrated over 7.8'' centered on the maximum continuum emission. 
The broken line shows the position of the fitted continuum.}
\label{fig:wrbumps}
\end{figure}

Such features are typical of WC stars (Smith 1968; Conti \& Massey 1989). 
Nevertheless, narrow nebular emission
lines can give non negligible contributions around 4650 \AA, such as 
CIII4650\AA; OII4651\AA; [FeIII]4658\AA; 
HeII4686\AA; 
%Ne[IV]4714\AA\; 
%HeI4713\AA\; 
%CIV4658\AA; 
[ArIV]4711\AA\,4740\AA\ 
and NIII4634,4640\AA . In order to evaluate their contribution to the bump 
around 4650 \AA \ hence the significance of this bump, 
we have used the photoionization models  produced by Stasi\~nska \& Leitherer 
(1996 hereafter SL96) for evolving
starbursts. 
No model can exactly  match
the observed strong emission features [OIII]5007\AA\, [OIII]4363\AA\ , 
[NeIII]3869\AA\ and [OII]3727\AA. We then used a model giving
a reasonable 
agreement with our data, using parameters as close as possible to that 
of IZw 18. 
The model we used (named "iiicikii" in SL96) corresponds to 4 Myrs for the burst in agreement 
with  evolutionary models predictions of CMH94 using the  $H\beta$ equivalent width.
The results from the model are given in column 4 of Table \ref{tab:mesures}.
The NIII4634,4640\AA\ lines are not given by the models.
 However, this doublet is absent in the spectrum of 
 SBS 0335-052 (Izotov et al., 1997), a starburst galaxy
 with an 
abundance and an electronic temperature very similar to those of IZw 18, and 
so the doublet was neglected. \\
 We then  measured the flux in the WR bump at 4650 \AA \  and  subtracted the 
 expected nebular lines given by the model.  
We find that the remaining flux in the bump at 4650 \AA \ is centered at
 around 4646 \AA , has a FWHM of 55 $\pm$ 5 \AA \ leading to a ratio 
WR(bump)/H$\beta$=0.029.
Finally, we have converted the measurements to absolute flux, assuming a 
distance of 10 Mpc for IZw 18. The flux in the bump at 4645 \AA \ is 
 $(1.0 \ \pm 0.3) \ 10^{37} \ $ ergs/s and, after subtraction of the 
nebular lines, $(9.90 \ \pm 3 ) \ 10^{36} \ $ ergs/s. \\
In the region around the HeI5876\AA\ line, a faint large bump centered at 5820 \AA \ is 
observed (fig. \ref{fig:wrbumps}) with a FWHM of 50 $\pm$ 10 \AA. 
The flux emitted in this bump is found to be $(4 \ \pm \ 1.5) \ 10^{36}  \ $ erg/s. \\
We also investigated the spatial location of the emission features. The nebular emission
is shifted by 1'' in the NE direction with respect to the continuum emission.
 By binning the 
spectrum over 1.6''  at all the positions along the slit, we find 
that the bumps at 
4645 \AA \ and 5820 \AA \ are correlated in position and occur in a region
 situated between 1'' and and 2'' 
SW from the central star cluster. Moreover, this corresponds to
 the position of the maximum 
emission of the narrow HeII4686\AA\ relative to H$\beta$
 (Fig. \ref{fig:correlation}). \\

\psfig{figure=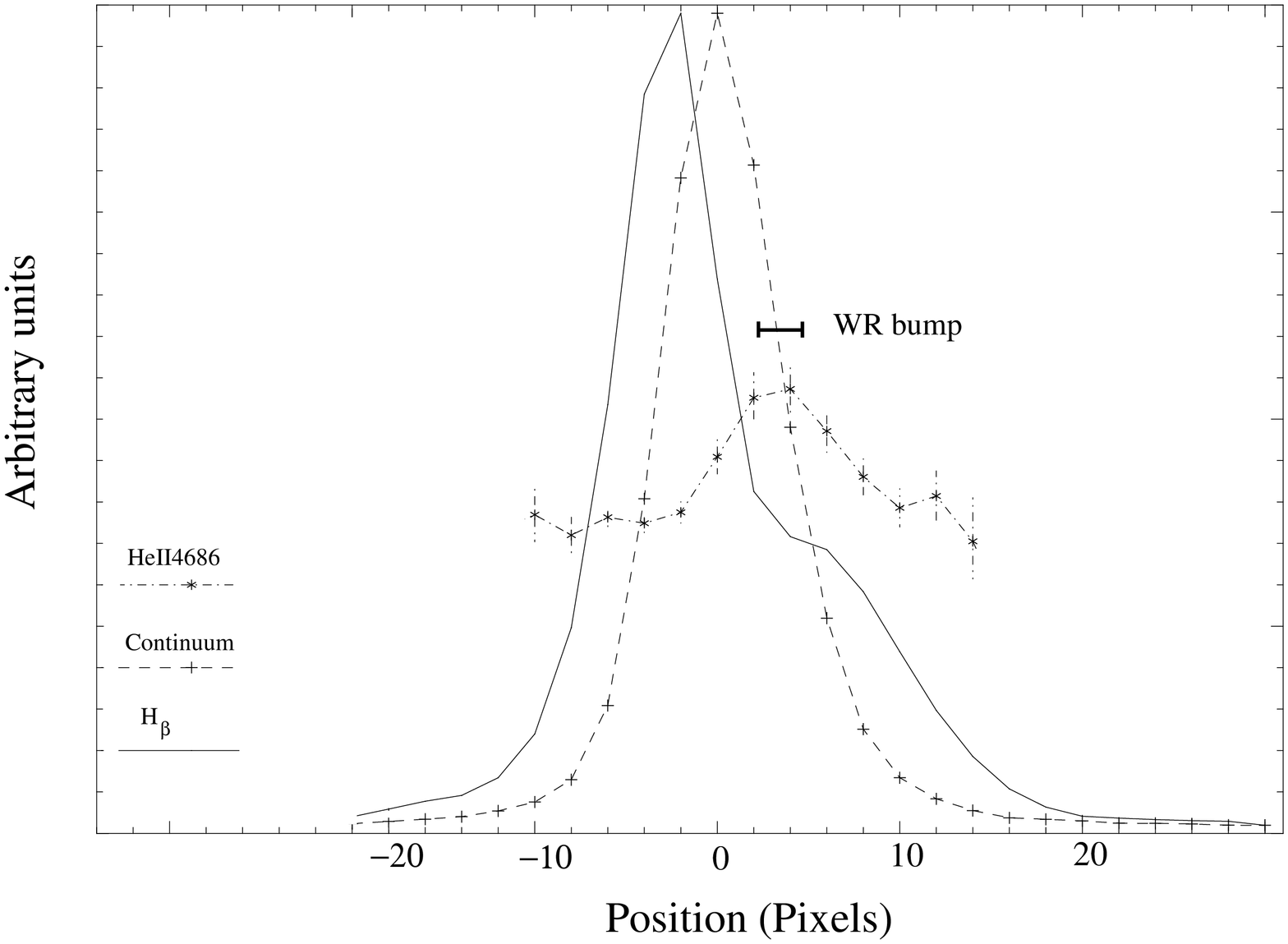,height=7cm,angle=-90}
\begin{figure}
%\picplace{7cm}
\caption[ ]{Spatial location of the different spectral emission features along the slit 
(1 pixel=0.3145 arcsec). Position of the maximum emission in the WRbump
 is indicated with bold line.}
\label{fig:correlation}
\end{figure}

\section{Discussion}
\subsection{WR population}
Generally WR stars are classified in two groups, WN and WC. Stars with type ranging from 
2 to 5 and from 4 to 6 respectively are called early (and noted WNE and WCE) while WR
with type ranging from 6 to 9 for WN and from 8 to 9 for WC are called late
 (Conti et al., 1990) and noted WNL, WCL. It is commonly accepted that early type
 stars are hotter than
late type ones, even if no direct relationship between type and temperature has been 
determined (Vacca \& Conti, 1992). WN stars are mainly characterized by the
NIII4634,4640\AA\ blend, NIV4057\AA\, NV4604,4620\AA\ blend 
and HeII4686\AA\ emission lines, while WC stars are betrayed by  CIII4645\AA, CIII5696\AA\,
 and
CIV5801,5812\AA\ emission features 
(Conti et al., 1990). \\
Our spectra show unambiguously a broad bump centered at 4645 \AA \ which cannot be nebular
in origin. No underlying contribution under the narrow 4686 \AA \ line is detected, which rules out
 WN stars for which a strong  broad contribution at 4686 \AA \ is expected,
 but not 
WCE in which the contribution at 4686 \AA \ can be negligible (Schaerer 1996).
The bump at 4645 \AA \ appears to be correlated with another bump around 5820 \AA \
which we interpret
as the CIV5801,5812\AA\ blend. 
Smith (1991) gives an average luminosity of $ \rm 5 \ 10^{36} \ ergs \ s^{-1} $ 
in the 4650 \AA \ bump and $ \rm 3 \ 10^{36} \ ergs \ s^{-1} $
in the 5808 \AA \ bump for WCs. The measured fluxes in the two bumps, taking the uncertainties into 
account, agree well with the presence of one or two WC stars in IZw 18. Note that  Hunter \& 
Thronson (1995)  report the possible detection of two WR stars with the HST using a 4695 \AA \ filter. 
 Finally, the comparison between the FWHM of the bumps given by Smith et al. (1990) with
our measured values  (55 $\pm$ 5 \AA \ at 4645 \AA \ and 50 $\pm$ 10 \AA \ at 5820 \AA) supports
 the presence of WCEs and indicates that the most probable types for these 
stars are WC4 or WC5. According to CMH94, and M95,
 the progenitors of WR 
stars at $Z = 0.001 $  are stars at least more massive than $\rm 80 \  M_{\odot}$. This
 constrains the IMF upper mass cutoff in IZw 18 to be higher than $\rm 80 \  M_{\odot}$. Moreover,
 at metallicity 
lower than 1/20th solar, CMH94 do not predict the formation of WC. However
these types are predicted if WR binary stars are taken into account (SV96; Cervi\~no et al. 1996) 
as will be discussed below. \\

\subsection{Narrow HeII4686\AA \ line and evolutionary models}
It is well known that IZw 18 presents a strong HeII4686\AA \ line in emission 
(SK93). The production of this line requires very energetic photons 
(E $\geq$ 54 eV) of which too few are produced  by ionizing sources 
with effective temperature 
$T_{eff} \ \leq$  70000 K (Garnett et al. 1991 hereafter G91). Since its intensity is several
 times larger 
than predicted by  photoionization models of HII regions ionised by O stars, Bergeron (1977) 
suggested that this line can arise directly in the atmosphere of hot Of stars. However,
 as asserted by Conti (1991), Of stars 
typically have both NIII4640\AA\ and HeII4686\AA\ with roughly the same
 intensity, and in IZw 18 no 
NIII4640\AA\ as strong as the HeII4686\AA\ line is observed. 
Campbell et al. (1986) have suggested 
that the low abundance in IZw 18 may suppress the NIII lines (see also Walborn et al 1995).
On the other hand, G91 have proposed an 
excitation of the HeII4686\AA\ by X-ray sources, but Motch et al. (1994) using ROSAT data, have shown 
that this mechanism cannot explain the observed emission in IZw 18. Pakull \& Motch (1989) have also 
suggested that hot WN 
stars could be at the origin of this line. Finally, ionization by WC stars has been suggested by
 Schaerer (1996). \\ 
Some association between HeII4686\AA\ and WO stars has been reported by G91 while nebular 
HeII4686\AA\ associated with the presence of WC stars have been reported by Gonzalez-Delgado et al. (1994).
The correlation observed between the maximum emission of the narrow nebular HeII4686\AA\ and the supposed 
location of the detected WC in IZw 18 (Fig. \ref{fig:correlation}) favour this
later hypothesis.  G91 however find no offset between the peaks of
 H$\beta$ and HeII4686\AA\ for a 
different orientation of the slit. Izotov \& Thuan (1997, ApJ submitted)  also report a shift
and attribute the 
difference between their results and the ones of G91 to a poorer S/N and resolution 
of the G91 data. \\
Schaerer (1996), using non-LTE, line blanketed model atmospheres  accounting for stellar
 winds, synthesized the 
nebular and WR HeII4686\AA\ emission in young starbursts. He finds that after 3 Myrs, the 
$\rm HeII_{nebular}/H\beta$  ratio increases due to the appearance of WC stars. For metallicities
 between solar 
and 1/5th solar, the ratio is the strongest with typical values between 0.01 and 0.03.
 At low metallicity 
($ 1/20 Z_{\odot}$), this ratio peaks after 3.4 Myrs at $4 \ 10^{-3}$, already ten times lower than
 what is observed in 
IZw 18. Moreover, at low $Z$,  due to the low mass loss, the WC population becomes negligible (M95, 
CMH94,  Maeder \& Meynet 1994)  explaining the faintness of the expected
 nebular 
HeII4686\AA\ line. \\
However, M95 has shown that models using  mass loss 
rates  twice the standard ones, although in good agreement with
 the overall results 
obtained by CMH94, predicts more WCs. Observational evidence for larger values 
of mass loss rates are given in Heap et al. (1994) for R136a.
 Still, some objects with relatively low metallicity 
exhibit large numbers of WC stars like  IC 10 which have a ratio
 WC/WN 
of 2 (Massey 1996) while Massey \& Armandroff (1995) suggest that the star formation ``vigor''
 affects the
IMF and thus the number of WC to WN stars (see eg M95).
Another way to form WR stars is the binary channel, but as 
mentioned by SV96 and
 Cervi\~no et al. (1996), the WRs formed in binary systems start to appear at 5 Myrs which may be longer 
than the burst age in IZw 18. 
Our new observations show that WC stars can form in a very metal deficient environment
 and tend to corroborate 
the high mass loss rate hypothesis of M95, possibly with rates even higher than
 twice the standard
one at very low metallicities. 
Although our result has little statistical bearing, the absence of WN stars comes as a surprise as 
evolutionary models 
(CMH94, M95) predict more WN stars than WC stars at low metallicity and 
even no WC at  $Z \leq \frac{1}{20} Z_{\odot}$ (CMH94). This detection of WC 
in a environment with metallicity as low as 1/40th solar may indicate that a 
binary channel for WR  star formation and/or higher mass loss rates have to be accounted for.

% +++++++++++++++++++++++++++++++++++++++++++++++++++++++++++++++++++++++

\section{Conclusion}
 
Two broad bumps have been detected in the spectra of IZw 18, centered respectively at 4645 \AA \ 
and 5820 \AA. We interpret these features as evidences for the presence of WR stars of WC type. 
The flux and FWHM of these bumps affirm that we are in presence 
of one or two WC4 or WC5 stars. The strong narrow HeII4686\AA\ line peaks 
co-spatially with the WR bumps indicating that this line is nebular in origin and due to
the presence of these detected WC stars. No evidence for the presence of WN stars is found contrary to
evolutionary models at very low metallicity.
This favours the hypothesis that mass loss rates may be higher than twice the standard one at
 very low metallicities or that the binary channel is an important process of 
WR stars formation. Finally,
the implication on the IMF of IZw 18 is that stars more massive than
$80 \ M_{\odot} $ have been formed in this galaxy.

% +++++++++++++++++++++++++++++++++++++++++++++++++++++++++++++++++++++++

\begin{acknowledgements}
We thank D. Schaerer and L. Drissen for helpful discussions and suggestions.
We thank the referee P.S. Conti for his helpful comments. 
JRR is funded by the Natural Sciences and Engineering
Research Council of Canada, and the Fonds FCAR of the
Government of Qu\'ebec.

\end{acknowledgements}

\end{document}